\documentclass[11pt]{article}

\title{Some Recent Progress in Classical General Relativity}
\author{Felix Finster,\ 
Joel Smoller\thanks{Research supported in part by the NSF, Grant No.\ 
DMS-G-9802370.}, and Shing-Tung Yau\thanks{Research supported in part 
by the NSF, Grant No.\ 33-585-7510-2-30.}}
\date{September 1999}

\newtheorem{Def}{Def.}[section]
\newtheorem{Thm}[Def]{Theorem}
\newtheorem{Prp}[Def]{Proposition}
\newtheorem{Lemma}[Def]{Lemma}
\newcommand{\Proof}{{\em{Proof: }}}
\newcommand{\QED}{\ \hfill $\FBox$ \\[1em]}

\newcommand{\spc}{\;\;\;\;\;\;\;\;\;\;}
\newcommand{\bra}{\mbox{$< \!\!$ \nolinebreak}}
\newcommand{\ket}{\mbox{\nolinebreak $>$}}

\newcommand{\R}{\mbox{\rm I \hspace{-.8 em} R}}
\newcommand{\1}{\mbox{\rm 1 \hspace{-1.05 em} 1}}

\newcommand{\sR}{\mbox{\rm \scriptsize I \hspace{-.8 em} R}}

\newcommand{\FBox}{\rule{2mm}{2.25mm}}

\begin{document}
\include{epsf}
\maketitle
In this short survey paper, we shall discuss certain recent results in 
classical gravity. Our main attention will be restricted to two topics 
in which we have been involved: the positive mass conjecture and its 
extensions to the case with horizons, including the Penrose 
conjecture (Part I), and the interaction of gravity with other 
force fields and quantum-mechanical particles (Part II).\\[1.5em]
{\Large{\bf{I. Positive Mass Conjecture and Related Topics}}} \\[1em]
One of the most difficult problems in classical relativity is to 
understand how and when singularities form. In the 1960's, Hawking and
Penrose proved that the existence of a closed trapped surface in an
asymptotically flat spacelike hypersurface gives rise to a singularity
in space-time. However, no proof based on pure partial differential equation 
arguments was found, and many questions remain unanswered.

Given an initial data set $(g_{ij}, p_{ij})$ on a three-dimensional 
manifold so that $g_{ij}$ is asymptotically Euclidean and $p_{ij}$ 
(the induced second fundamental form in an embedding)
falls off asymptotically, it is interesting to ask the following 
questions.
\begin{enumerate}
\item When will such an initial data set contain a closed trapped surface? 
If so, how to locate it?
\item If the initial data contains no closed trapped surface, how to 
tell whether such a surface will appear at a later time under the 
evolution of Einstein's equations?
\item If we assume that the trace of $p_{ij}$ is zero and that under 
the evolution this condition is preserved, will a singularity occur 
without the existence of a closed trapped surface for all time?
\item If a singularity does occur, what is the structure of the null 
geodesics in a neighborhood of the singularity, and what is the 
structure of the curvature tensor in this neighborhood? What is the 
criterium on the initial data set for the curvature to blow up at the 
singularity?
\item Can one define physically relevant local (or quasi-local) 
quantities such as mass and angular momentum to describe regions in a 
strongly gravitationally interacting space-time? For example, when two 
bodies interact, what is the binding energy and what is the mass of 
the resulting configuration? How can one estimate the gravitational 
radiation for strongly interacting bodies? How can one justify
the linearized theory of gravitational radiation?
\end{enumerate}
For all the above questions related to singularity formation, one
usually studies only generic initial data. However, it has 
been a difficult problem in nonlinear partial differential equations 
to understand how to perturb away the singularity.

For all these questions, it would be good if the known class of 
spherically symmetric solutions of the Einstein equations were rich. 
Except for the Schwarzschild case, such solutions cannot be
vacuum solutions. Hence to consider these questions, one is forced
to couple gravity to other matter fields. For 
the case of a massless scalar field, Christodoulou (\cite{C1, C2, C3, C4})
has studied the question of the 
formation of singularities quite extensively. If a singularity exists, it is 
located at the origin. While much is known in this 
case, the details of how the singularity forms is still poorly 
understood. (When naked singularities form, one would like to know the 
behavior of null geodesics.) Based on numerical studies, Choptuik 
\cite{Ch} has 
found the new phenomena that, for a one-parameter family of initial data, 
the mass function exhibits some critical phenomena similar to those which
occur in statistical mechanics, at the time when a black hole forms.
However, a detailed theoretical study is lacking.
An interesting consequence of the above study is that after 
gravitational radiation, the space-time is either time-asymptotic to the 
flat space-time or to the Schwarzschild space-time. This raises an 
interesting question when we couple gravity to a Yang-Mills 
field or to Dirac spinors: what is the possible asymptotic state of 
spherically symmetric initial data? Would those stable coupled solutions found
by us (cf.\ Section II below) be the only possible states?
When we want to extend the spherically symmetric case to an 
axisymmetric geometry, the space-time is far more complicated. While 
it is clear that angular momentum may be used to make many 
configurations stable, the number of degrees of freedom is large and it is 
difficult to find solutions of gravity coupled to other 
fields. (For stationary black holes with a vacuum background, it has 
to be the Kerr solution.) It is still not known whether one can find 
multi-black holes which can be stabilized by the addition of angular momentum.

Beyond axisymmetric solutions, Bartnik \cite{Bn1, Bn2} proposed a class of 
initial data sets which can be foliated by round spheres. Using this 
ansatz, he was able to parametrize a large set of initial data
having zero or nonnegative scalar curvature. (For initial data set, if it 
is a maximal slice, the scalar curvature is always nonnegative.)
According to his numerical study, this ansatz has been very useful in 
understanding radiation of a single black hole. Perhaps the 
theoretical study for critical data in this class would be interesting.

Let us now turn to general space-time with no spherical symmetry. We 
restrict ourselves to asymptotically flat space-times. In this case, 
we have asymptotic space-time Lorentzian symmetry. Based on this asymptotic 
symmetry, it is well known that one can define the concept of mass 
and linear momentum associated to each initial data set (which is 
invariant under the Lorentzian symmetry at asymptotic infinity 
\cite{asymm}).
About twenty years ago, Schoen-Yau \cite{SY} (subsequently \cite{W} 
and others) proved the positive mass conjecture which says that the 
total (mass, linear momentum) is a non-spacelike four-vector.
The total mass is therefore always nonnegative. It is zero only when 
the space-time is flat.

The positivity of the mass says that the trivial space-time is stable 
(the dynamic stability among a class of reasonable initial data has 
recently been demonstrated by Christodoulou and Klainerman \cite{CK}).
However, the nonlinear stability of the Schwarzschild solution is still 
unknown. Based on the ``Cosmic Censorship conjecture,'' Penrose 
proposed an inequality relating the total 
mass of the black hole to the area of the outermost horizon. It 
says that among all initial data sets with fixed mass, the 
time-symmetric Schwarzschild solution initial data set has the 
largest area for its outermost apparent horizon.

In general, if the initial data set is a maximal slice for the 
space-time, the scalar curvature of the three-dimensional manifold is 
non-negative. In such a case, the conjecture of Penrose was recently 
settled by Huisken-Ilmanen \cite{HI}, obtaining the optimal result
only under the assumption that the 
outermost black hole is connected. It was based on an idea of Geroch 
that the Hawking (quasi-local) mass is monotonic along an evolution of 
a surfaces $\Sigma_t$ which starts from the hole to the sphere at 
infinity. The evolution is governed by the requirement that it moves 
the surfaces along the normal direction and with magnitude minus the inverse of 
mean curvature. Geroch notices that for the sphere at inifinity, the 
Hawking mass is simply the total mass of the initial data set while at 
the black hole, the Hawking mass is, up to a universal constant, the 
square root of the area of the black hole. Hence if the flow of the 
surface $\Sigma_t$ exists, the Penrose conjecture would then be proved.
Huisken and Ilmanen developed the mathematical framework in which 
these ideas could be made precise. However, the flow exhibits
jump phenomena and much care is needed to assure that the 
inequality jumps in the right manner. Much more recently, H.\ Bray 
\cite{B} has been able to improve the result in the case of a non-connected
outermost horizon by a new method, partly relying on the ideas of 
Schoen-Yau \cite{SY} and the curvature estimates \cite{BF}.

For the proof of the Penrose conjecture, one still must answer 
the question as to when the initial data set is not maximal.
It would also be nice to see the corresponding inequality for the 
Bondi mass (total mass after radiation).

Besides the total mass and linear momentum, an important conserved 
quantity is angular momentum. This was studied extensively by Ashtekar 
\cite{A}. One needs to study the relation between angular momentum and other 
conserved quantities such as the total mass and linear momentum. It 
seems reasonable to believe that the total mass should dominate the 
square of the angular momentum if the initial data set is nonsingular.

To better understand angular momentum, Huisken and Yau \cite{HY} defined the
concept of center of mass of an initial data set. 
It is Lorentzian invariant and, remarkably, under the evolution of 
Einstein's equations, the velocity of the center of gravity is the 
linear momentum divided by $2m$, where $m$ is the total mass of the 
initial data set.

One hopes to study all possible naturally conserved quantities and the 
relations among these conserved quantities, when the initial data set 
is non-singular. It is always interesting to know how radiation 
effects all those quantities. For an isolated gravitational system, 
what would be its asymptotic state after radiation? We conjecture that
the time-asymptotic state is just the superposition of several known
stationary solutions including the charged Kerr black holes and the static
coupled solutions found above (e.g.\ when we are coupling the Einstein equations
to the Yang-Mills, Dirac particles, or a real scalar field).

The global behavior of the Einstein system is difficult to study, 
partially because we do not have (quasi-)local quantities which 
behave well under time evolution. The Hawking mass is one such 
example. It is monotonic in some directions. Unfortunately, it is 
not positive in general. For certain important closed surfaces, which 
are obtained by minimizing area under a volume constraint, 
Christodoulou and Yau were able to prove the positivity of the Hawking 
mass \cite{CY}. However, they assumed that the scalar curvature is 
non-negative. It would be nice to remove this assumption.

If one considers sufficient conditions for the formation of black holes
in a general setting, the best theorem is due to Schoen-Yau \cite{SY2}.
This says that, by suitably defining the diameter $d(\Omega)$ of a region
$\Omega$, then if the matter density in the region $\Omega$ is greater 
than $d^{-2}$ up to a universal constant, a closed trapped 
surface can be found. This implies that a black-hole type singularity exists.
In this theorem, the existence of black holes results from the 
condensation only of matter. It would be desirable to include the 
contribution of gravitation effects. Namely, it is interesting that in the 
argument by Shoen-Yau, only the lower bound of the first eigenvalue of the 
operator $-\Delta + \frac{1}{6} \:R$ is used.
In the time symmetric case (i.e. with $p_{ij} \equiv 0$),
$\frac{1}{2}\:R$ is the local matter density. It would be nice to see 
if this method can be extended to the general case, in the sense that
the spectrum of some operator can be used to yield a condition for
the formation of black holes.\\[1.5em]
{\Large{\bf{II. The Interaction of Gravity with Other Force 
Fields and Dirac Particles}}} \\[1em]
According to Einstein's Theory of General Relativity, gravity 
is described geometrically through Einstein's equations. The 
understanding of gravity has been driven by the discovery of special 
solutions of these equations. The most important examples are
the Schwarzschild solution, the Kerr-Newman solution, and the
Friedmann-Robertson-Walker solution \cite{ABS}. Particularly 
interesting effects are obtained when one couples gravity,
as expressed through Einstein's equations, to other 
fundamental force fields. The simplest such example is the 
Reissner-Nordstr\"om solution resulting from the coupling of gravity 
to electromagnetism (Maxwell's equations). This solution, like the 
Schwarzschild solution, has an essential singularity at the 
origin. The generalization to non-abelian Yang-Mills fields led to 
the discovery of Bartnik and McKinnon (BM) \cite{BM} (see also~\cite{KM, SW})
of everywhere regular solutions.
This came as a surprise because several results for related systems led to
the conjecture that such solutions cannot exist. Indeed, neither the
vacuum Einstein equations, nor the pure Yang/Mills equations have non-trivial
static, globally defined, regular solutions~\cite{C, D}. The existence 
of these solutions depends on the coupling of the different force 
fields, whereby the attractive gravitational force is balanced by the 
YM repulsive force. But this balance is rather delicate; for example, 
the BM solutions are unstable with respect to small perturbations
\cite{SZ}.
Other interesting solutions of Einstein's equations result from 
coupling gravity to quantum mechanical matter fields. The case of a 
complex scalar field was considered by T.D.\ Lee et al \cite{L}, who 
found soliton-like solutions modeling (bosonic) stars; see too
D.\ Christodoulou \cite{C2} who studied the gravitational collapse of a 
real massless scalar field.

We report here on recent work (see~\cite{FSY1}--\cite{FSY4} for 
details) of a different type of coupling; namely, gravity 
coupled to both quantum mechanical particles with spin (Dirac 
particles), and to an electromagnetic field. We first study the 
resulting Einstein-Dirac-Maxwell (EDM) equations for a static, 
spherically symmetric system of two fermions in a singlet spinor 
state. We find stable soliton-like solutions, and we discuss 
their properties for different values of the electromagnetic 
coupling constant. We note too that the inclusion of gravity has a 
regularizing effect on solutions, in the sense that our solutions are 
more regular than one would expect from a naive analysis of the
Feynman diagrams; see \cite{FSY5}. We then study 
black-hole solutions for these equations (see \cite{KM, SWY}),
and we find, surprisingly,
that under rather weak regularity conditions on the form of the event 
horizon, the only black-hole solutions of the EDM equations are the 
Reissner-Nordstr\"om (RN) solutions. That is, the spinors must vanish 
identically. Applying this to the gravitational collapse to a black 
hole of a ``cloud''of relativistic spin-$\frac{1}{2}$-particles, our result 
indicates that the Dirac particles must eventually disappear inside 
the event horizon. We also show that the Dirac equation has no 
normalizable, time-periodic solutions in a RN black-hole background. 
The physical interpretation of this result is that the Dirac particles
cannot remain on a periodic orbit
around the black-hole. This result has recently been extended to an
axisymmetric black hole geometry~\cite{FKSY}.

In our study of the coupled EDM equations, we employ a special ansatz 
for the spinors. In this ansatz, we do not assume that the Dirac particles
are in a spherically symmetric state; indeed they are allowed to have angular
momentum. 
However, we arrange $(2j+1)$ of these particles in such a way that the 
total system is static and spherically symmetric. (In the language of 
atomic physics, we consider the completely filled shell of states 
with angular momentum $j$. Classically, this multi-particle system can 
be thought of as several Dirac particles rotating around a common 
center such that this angular momentum adds up to zero.) Since the 
system of fermions is spherically symmetric, we obtain a consistent 
set of equations if we also assume spherical symmetry for the 
gravitational and electric fields. We can thus separate out the 
angular dependence, and the problem then reduces to a system of 
nonlinear ODEs.

\section{The EDM Equations}
\label{sec2}
The general Einstein-Dirac-Maxwell equations are
\begin{equation}
R^i_j \:-\:\frac{1}{2}\:R\:\delta^i_j \;=\; -8 \pi \:T^i_j \;\;\;,\;\;\;\;\;
(G-m) \:\Psi_a \;=\; 0 \;\;\;,\;\;\;\;\; \nabla_k F^{jk} \;=\;
4 \pi e \sum_a \overline{\Psi_a} G^j \Psi_a \;\;\;,
        \label{eq:21}
\end{equation}
where $T^i_j$ is the sum of the energy-momentum tensor of the Dirac 
particles and the Maxwell stress-energy tensor. The $G^j$ are the 
Dirac matrices which are related to the Lorentzian metric via the 
anti-commutation relations
\[ g^{jk}(x) \:\1 \;=\; \frac{1}{2}
\left\{ G^j(x), G^k(x) \right\} \;\equiv\; \frac{1}{2} \:
(G^j G^k + G^k G^j)(x) \;\;\;. \]
$F^{jk}$ denotes the electromagnetic field tensor, and $\Psi_a$ are 
the wave functions of fermions of mass $m$ and charge $e$. The Dirac 
operator is denoted by $G$, and it depends on both the gravitational 
and electromagnetic field; for details see \cite{FSY1, FSY2}.

We now specialize to the case of static, spherically symmetric 
solutions of the EDM system (\ref{eq:21}). In polar coordinates $(t, 
r, \vartheta, \varphi)$, we write the metric in the form
\begin{equation}
ds^2 \;=\; \frac{1}{T(r)^2} \:-\: \frac{1}{A(r)} \:dr^2 \:-\: r^2 
\:(d\vartheta^2 + \sin^2 \vartheta \:d\varphi^2)
        \label{eq:21a}
\end{equation}
with positive functions $T$ and $A$.
Depending on whether we consider particlelike solutions or 
black-hole solutions, the region of space-time which we consider is 
$r>0$, or $r>\bar{r}>0$, respectively; in the latter case, we assume 
that $r=\bar{r}$ is the event horizon. We always consider solutions for which 
the metric (\ref{eq:21a}) is Minkowskian,
\begin{equation}
\lim_{r \rightarrow \infty} A(r) \;=\; 1 \;=\; \lim_{r \rightarrow 
\infty} T(r) \;\;\;,
        \label{eq:22}
\end{equation}
and has finite (ADM) mass; i.e.,
\begin{equation}
\lim_{r \rightarrow \infty} \frac{r}{2} \:(1-A(r)) \;=\; \rho \;<\; 
\infty \;\;\;.
        \label{eq:23}
\end{equation}
In the static case, the fermions only generate an electric field, and 
thus we may assume that the electromagnetic potential ${\cal{A}}$ has the 
form ${\cal{A}}=(-\phi, \vec{0})$, where $\phi=\phi(r)$ is the Coulomb 
potential.

The Dirac operator $G$ can be written as
\begin{eqnarray}
\lefteqn{ G \;=\; i G^j(x) \:\frac{\partial}{\partial x^j} \:+\: B(x) }
\nonumber \\
&=& i T \gamma^0 \left(\frac{\partial}{\partial t} - i e \phi \right)
        + \gamma^r \left( i \sqrt{A} \frac{\partial}{\partial r} + \frac{i}{r} 
        \:(\sqrt{A}-1) -\frac{i}{2}\:\sqrt{A} \:\frac{T^\prime}{T} \right)
        + i \gamma^\vartheta \frac{\partial}{\partial \vartheta} + i
        \gamma^\varphi \frac{\partial}{\partial \varphi} \; , \;\;\;\;\;\;
        \label{eq:24}
\end{eqnarray}
where $\gamma^t$, $\gamma^r$, $\gamma^\vartheta$, and 
$\gamma^\varphi$ are the $\gamma$-matrices in polar coordinates, in 
Minkowski space; namely
\begin{eqnarray*}
\gamma^t &=& \gamma^0 \\
\gamma^r &=& \gamma^1 \:\cos \vartheta \:+\: \gamma^2 \:
        \sin \vartheta \: \cos \varphi \:+\: \gamma^3 \: \sin \vartheta \:
        \sin \varphi \\
\gamma^\vartheta &=& \frac{1}{r} \left( -\gamma^1 \:\sin \vartheta \:+\: \gamma^2 \:
        \cos \vartheta \: \cos \varphi \:+\: \gamma^3 \: \cos \vartheta \:
        \sin \varphi \right) \label{2.13aa} \\
\gamma^\varphi &=& \frac{1}{r \sin \vartheta} \left( -\gamma^2 \: \sin \varphi \:+\:
        \gamma^3 \: \cos \varphi \right) \spc ,
\end{eqnarray*}
where
\[      \gamma^0 \;=\; \left( \begin{array}{cc} \1 & 0 \\ 0 & -\1 
        \end{array} \right) \;\;\;,\spc \gamma^i \;=\;
        \left( \begin{array}{cc} 0 & \sigma^i \\ -\sigma^i & 0
        \end{array} \right) \;\;,\;\;\; i=1,2,3, \]
and $\sigma^i$ denote the Pauli matrices.

In analogy with the central force problem in Minkowski space \cite{S}, 
this Dirac operator commutes with: a) the time translation operator 
$i \partial_t$, b) the total angular momentum operator $J^2$, c) the 
$z$ component of the total angular momentum $J_z$, and d) with the 
operator $\gamma^0 P$, where $P$ is the parity. Since these operators 
also commute with each other, any solution of the Dirac equation can 
be written as a linear combination of solutions which are simultaneous 
eigenstates of these operators. We use this ``eigenvector basis'' to 
separate out both the angular and time dependence, and to calculate 
the total current and energy momentum tensor of the Dirac particles. 
Using the ansatz in \cite{FSY1, FSY2, FSY3}, we can describe the Dirac 
spinors using two real functions $\alpha, \beta$. We arrive at 
the following system of ordinary differential equations for the $5$ 
real functions $\alpha$, $\beta$, $A$, $T$, and $\phi$:
\begin{eqnarray}
\sqrt{A} \:\alpha^\prime &=& \pm \frac{2j+1}{2 r} \:\alpha \:-\: 
        ((\omega-e \phi) T + m) \:\beta \label{eq:E1} \\
\sqrt{A} \:\beta^\prime  &=& ((\omega-e \phi) T - m) \:\alpha \:\mp\:
        \frac{2j+1}{2 r} \:\beta \label{eq:E2} \\
r \:A^\prime & = & 1-A \:-\: 2(2j+1) (\omega-e \phi) T^2
        \:(\alpha^2 + \beta^2) \:-\: r^2 A T^2\: |\phi^\prime|^2
        \label{eq:E3} \\
2 r A \:\frac{T^\prime}{T} &=& A-1 \:-\: 2(2j+1) (\omega-e \phi) T^2 
        \:(\alpha^2+\beta^2) \:\pm\: 2 \:\frac{(2j+1)^2}{r} \:T\: \alpha \beta 
        \nonumber \\
&&+2(2j+1) \:m T \:(\alpha^2-\beta^2) \:+\: r^2 A T^2\: |\phi^\prime|^2
        \label{eq:E4} \\
r^2 A \:\phi^{\prime \prime} &=& -(2j+1)\: e \:(\alpha^2+\beta^2)
        -  \left( 2r A  + r^2 A \:\frac{T^\prime}{T}
        + \frac{r^2}{2} \:A^\prime \right) \phi^\prime \;\;\;\; .
        \label{eq:E5}
\end{eqnarray}
Equations (\ref{eq:E1}) and (\ref{eq:E2}) are the Dirac equations 
(the $\pm$ signs correspond to the two possible eigenvalues of 
$\gamma^0 P$); (\ref{eq:E3}) and (\ref{eq:E4}) are the 
Einstein equations, while Maxwell's equations reduce to the single 
equation (\ref{eq:E5}). Here $j=\frac{1}{2}, \frac{3}{2},\ldots$, the 
constant $\omega$ enters via the plane wave dependence of the spinors; 
namely $\exp(-i \omega t)$, and as for the general equations 
(\ref{eq:21}), $m$ and $e$ denote the mass and charge, respectively, 
of the fermions. We also require that, in addition to (\ref{eq:22}), 
(\ref{eq:23}), the electromagnetic potential vanishes at infinity,
\begin{equation}
\lim_{r \rightarrow \infty} \phi(r) \;=\; 0 \;\;\;.
        \label{eq:210}
\end{equation}
Since the equations (\ref{eq:E1})--(\ref{eq:E5}) are invariant under 
the gauge transformations
\begin{equation}
\phi(r) \;\rightarrow\; \phi(r) + \kappa \;\;\;,\spc \omega 
\;\rightarrow\; \omega + e \kappa\;\;\;,\spc \kappa \in \R,
        \label{eq:211}
\end{equation}
we see that (\ref{eq:210}) can be fulfilled by a suitable gauge 
transformation, provided that $\phi$ has a limit at infinity.

In Sections \ref{sec2}-\ref{sec5}, we shall be concerned with two 
different types of solutions of equations 
(\ref{eq:E1})--(\ref{eq:E5}); namely {\em{particlelike solutions}} 
(smooth solutions defined for all $r \geq 0$), and {\em{black hole 
solutions}} (solutions defined for all $r>\bar{r}>0$, where 
$A(\bar{r})=0$ and $A(r)>0$ for all $r>\bar{r}$; $r=\bar{r}$ is the 
event horizon). In the first case, we require the following 
normalization condition on the spinors:
\begin{equation}
\int_0^\infty (\alpha^2 + \beta^2) \:\frac{T}{\sqrt{A}} \:dr \;=\; 
1 \;\;\;, \spc {\mbox{(particlelike)}}
        \label{eq:212}
\end{equation}
while in the second case we require that for all $r_0 > \bar{r}$,
\begin{equation}
0 \;<\; \int_{r_0}^\infty (\alpha^2 + \beta^2) \:\frac{T}{\sqrt{A}} \:dr 
\;<\; \infty \;\;\;\;\spc {\mbox{(black holes)}}.
        \label{eq:213}
\end{equation}
These conditions are necessary in order that the Dirac spinors define 
physically meaningful wave functions.

\section{Particlelike Solutions}
In this section we shall describe our numerical construction of 
particlelike solutions for the equations~(\ref{eq:E1})--(\ref{eq:E5}).
For simplicity we shall restrict ourselves to the case $j=1/2$.
We shall also discuss the stability and properties of the ground state 
solutions for different values of the electromagnetic coupling 
constant $(e/m)^2$. We shall show that solutions exist even when the 
em coupling is so strong that the total interaction is repulsive in 
the non-relativistic limit. In addition, for small em coupling, 
$(e/m)^2<1$, we shall show that {\em{stable}} particlelike solutions 
exist for small values of $m$, and using certain topological 
techniques, we show that this stable solution becomes unstable as $m$ 
increases.

The construction of particlelike solutions is obtained via a 
rescaling argument (see \cite{FSY1, FSY2}). The idea is to weaken the 
conditions (\ref{eq:21a}), (\ref{eq:210}), and (\ref{eq:212}) to
\begin{equation}
0 \;\neq\; \int_0^\infty (\alpha^2 + \beta^2) \:\frac{T}{\sqrt{A}} 
\:dr \;<\; \infty \;\;\;,\;\;\;\;\;
0 \;\neq\; \lim_{r \rightarrow \infty} T(r) \;<\; \infty \;\;\;,\;\;\;\;\;
\lim_{r \rightarrow \infty} \phi(r) \;<\; \infty \;\;\;,
        \label{eq:31}
\end{equation}
and instead set
\begin{equation}
T(0) \;=\; 1 \;\;\;,\spc \phi(0) \;=\; 0  \;\;\;,\spc m \;=\; 1 \;\;\;.
        \label{eq:32}
\end{equation}
This enables us to use a Taylor expansion around $r=0$, and we obtain 
the following expansions near $r=0$:
\[ \begin{array}{rclcrcl}
\alpha(r) &=& \alpha_1 \:r \:+\: {\cal{O}}(r^2) &\;\;\;,\;\;\;\;\;& 
\beta(r) &=& {\cal{O}}(r^2) \\
A(r) &=& 1 + {\cal{O}}(r^2) &\;\;\;,\;\;\;\;\;& 
T(r) &=& 1 + {\cal{O}}(r^2) \;\;\;,\;\;\;\;\;
\phi(r) \;=\; {\cal{O}}(r^2) \;\;\;.
\end{array} \]
Solutions to our equations now depend on the three real parameters 
$e$, $\omega$, and $\alpha_1$. For a given value of these parameters, 
we can construct initial data at $r=0$, and using the standard 
Mathematica ODE solver, we ``shoot'' for numerical solutions of the 
modified system~(\ref{eq:E1})--(\ref{eq:E5}), (\ref{eq:32}). By 
varying $\omega$ (for fixed $e$ and $\alpha_1$), we can arrange that 
the spinors $(\alpha, \beta)$ tend to the origin for large $r$, and 
the conditions (\ref{eq:23}) and (\ref{eq:31}) also hold.

Given a solution $(\alpha, \beta, A, T, \phi)$ of this modified 
system, we consider the scaled functions
\[ \begin{array}{rclcrcl}
\tilde{\alpha}(r) &=& \sqrt{\frac{\tau}{\lambda}} \:\alpha(\lambda r)
&\;\;\;,\;\;\;\;\;& 
\tilde{\beta}(r) &=& \sqrt{\frac{\tau}{\lambda}} \:\beta(\lambda r) \\
\tilde{A}(r) &=& A(\lambda r) &\;\;\;,\;\;\;\;\;& 
\tilde{T}(r) &=& \tau^{-1} \:T(\lambda r) \;\;\;,\;\;\;\;\;
\tilde{\phi}(r) \;=\; \tau \:\phi(\lambda r) \;\;\;.
\end{array} \]
By direct computation, these functions satisfy the original
equations~(\ref{eq:E1})--(\ref{eq:E5}) and the equations 
(\ref{eq:22}), (\ref{eq:23}), and (\ref{eq:212}), provided that the 
physical parameters are transformed according to
\begin{equation}
\tilde{m} \;=\; \lambda\:m \;\;\;,\spc \tilde{\omega} \;=\; \lambda 
\tau \:\omega \;\;\;,\spc \tilde{e} \;=\; \lambda \:e \;\;\;,
        \label{eq:32a}
\end{equation}
where the scale factors $\lambda$ and $\tau$ are given by
\[ \lambda \;=\; \left( 4 \pi \int_0^\infty (\alpha^2+\beta^2) 
\:\frac{T}{\sqrt{A}} \:dr \right)^{\frac{1}{2}} \;\;\;,\spc
\tau \;=\; \lim_{r \rightarrow \infty} T(r) \;\;\;. \]
Finally, the condition (\ref{eq:210}) can be fulfilled by a suitable gauge 
transformation. Notice that the parameter $(\tilde{e}/\tilde{m})^2 = e^2$ is 
invariant under the above scaling. It is thus convenient to table
$(\tilde{e}/\tilde{m})^2$ (and not $\tilde{e}^2$) as the parameter 
used to describe the strength of the em coupling. We point out that 
the above scaling technique is used only to simplify the numerics; 
for the physical interpretation, however, we must always work with 
the scaled (tilde) solutions. Since the transformation from the 
un-tilde to the tilde variables is one-to-one, our scaling method 
yields all the solutions of the original system. From now on, we shall 
only consider the scaled solutions, and for simplicity in notation, we 
shall omit the tilde.

\section{Properties of the Particlelike Solutions}
\begin{figure}[tb]
        \epsfxsize=12cm
        \centerline{\epsfbox{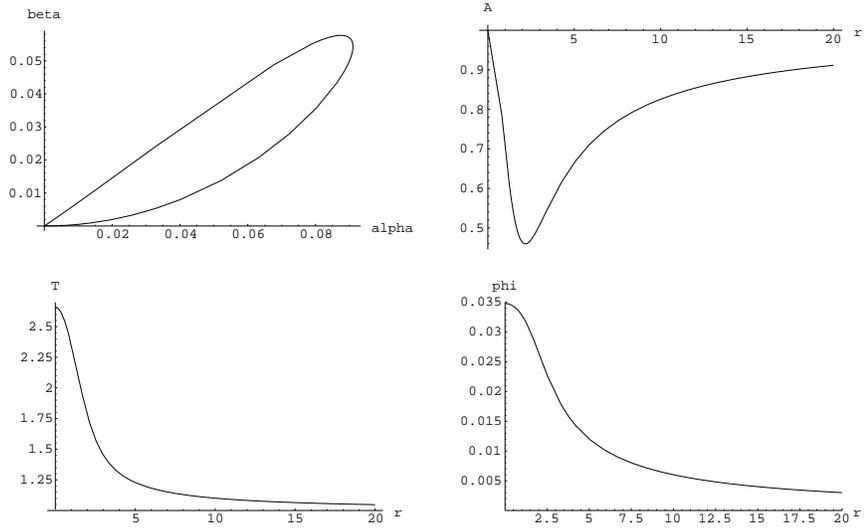}}
        \caption{Solution of the EDM equations for parameter values
        $(e/m)^2=0.7162$, $m=0.7639$, $\omega=0.6807$, $\rho=1.15416$
        ($\alpha^\prime(0)=0.05361$).}
        \label{fig1}
\end{figure}

\begin{figure}[tb]
        \epsfxsize=13cm
        \centerline{\epsfbox{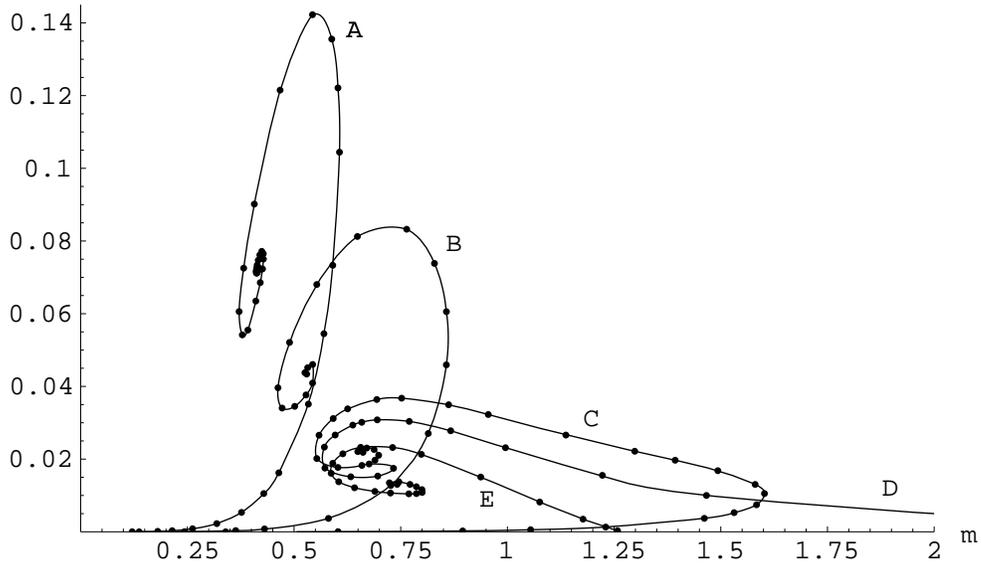}}
        \caption{Binding Energy $m-\omega$ of the Fermions for $(e/m)^2=0$ 
        (A), $0.7162$ (B), $0.9748$ (C), $1$ (D), and $1.0313$ (E).}
        \label{fig2}
\end{figure}

We have found solutions having different rotation numbers 
$n=0,1,2,\ldots$ of the vector $(\alpha, \beta)$. In the 
non-relativistic limit, $n$ is the number of zeros of the 
corresponding Schr\"odinger wave functions, and thus $n=0$ 
corresponds to the ground state, $n=1$ to the first excited state, 
and so on. However due to the nonlinearity of our equations, $n$ no 
longer has this simple interpretation. For simplicity in what 
follows, we shall only discuss the $n=0$ solutions. The 
graphs of a typical such solution is shown in Figure~\ref{fig1}.
For each solution, the spinors $(\alpha, \beta)$ decay exponentially to 
zero at infinity. We interpret this to mean that the fermions have a 
high probability to be confined to a neighborhood of the origin. In 
view of this rapid decay of the spinors, our solutions asymptotically 
go over into the spherically symmetric RN solutions of the 
Einstein-Maxwell equations \cite{ABS}, as $r \rightarrow \infty$. That 
is, for large $r$,
\[ A(r) \;\approx\; T^{-2}(r) \;\approx\; 1 - \frac{2 \rho}{r} + 
\frac{(2e)^2}{r^2} \;\;\; . \]
In other words, our solution, for large $r$, looks like the 
gravitational and electrostatic field generated by a point particle 
at the origin with mass $m$ and charge $2e$. Note that in contrast to 
the RN solution, however, our solutions have no event horizon or 
singularities. One can understand this from the fact that we consider 
here quantum mechanical particles, rather than point particles. 
Therefore the wave functions are de-localized according to the 
Heisenberg Uncertainty Principle, and so the distributions of matter 
and charge are also de-localized, thereby preventing the metric from 
forming singularities. In general, we can parametrize solutions by 
the rest mass $m$, and the energy $\omega$ of the fermions. In 
Figure~\ref{fig2},
we plot the binding energy $m-\omega$ versus $m$ 
for different values of the parameter $(e/m)^2$, and we see that 
$m-\omega$ is always positive, indicating that the fermions are in a 
bound state. For weak em coupling, $(e/m)^2<1$, the curve is a spiral 
which starts at the origin. The binding energy decreases for fixed 
$m$ and increasing $(e/m)^2$, since the em repulsion weakens the 
binding. The mass energy spectrum when $(e/m)^2 \ll 1$ becomes 
similar to the case of the Einstein-Dirac equations (without the em 
interaction); see~\cite{FSY1}. We can use linearization techniques to 
show numerically that for small $m$, if $(e/m)^2<1$, the solutions 
are stable with respect to spherically symmetric perturbations. For 
larger values of $m$, we can investigate the stability using Conley 
index theory (see~\cite{Sm}), where $m$ is taken to be the 
bifurcation parameter. This technique shows that the
stability/instability of a solution remains unchanged if $m$ is varied 
continuously and no bifurcations occur. Moreover, at bifurcation 
points, the Conley index theory provides a powerful technique to 
analyze changes of stability. Using this, we find that all solutions 
on the ``lower branch'' of the spiral curves A and B of 
Figure~\ref{fig2} (i.e., on the curve from the origin up to the 
maximal value of $m$), are stable, and all solutions on the ``upper 
branch'' are unstable.

From Figure~\ref{fig2}, we see that this form of the mass energy 
spectrum changes when $(e/m)^2 \approx 1$, the regime where, in the 
classical limit, the electrostatic and gravitational forces balance 
each other. To better understand this situation, we take the 
non-relativistic limit in our EDM equations. To do this, we fix 
$(e/m)^2$, and assume that $e$ and $m$ are small. In this limit, the 
coupling of the spinors to both the gravitational and em forces 
becomes weak: $A, T \approx 1$ and $\phi \approx 0$. The Dirac 
equations imply that $\omega \approx m$ and $\alpha \gg \beta$. Thus 
the EDM equations go over to the Schr\"odinger equation with the 
Newtonian and Coulomb potentials; namely,
\begin{eqnarray}
\left( -\frac{1}{2m} \:\Delta \:+\: e \phi \:+\: m V \right) 
\Psi &=& E \:\Psi \label{eq:33} \\
- \Delta V \;=\; -8 \pi \:m \:|\Psi|^2 \;\;\;,\spc
- \Delta \phi &=& 8 \pi \:e \:|\Psi|^2 \;\;\;, \label{eq:34}
\end{eqnarray}
where $E=\omega - m$, $\Psi(r)=\alpha(r)/r$, $V(r)=1-T(r)$, and 
$\Delta$ is the radial Laplacian on $\R^3$. From (\ref{eq:34}) we see 
that the Newtonian and Coulomb potentials are multiples of each other; 
namely $V=-m/e \:\phi$. Thus if $(e/m)^2 \geq 1$, the total interaction 
is repulsive so that the Schr\"odinger equation (\ref{eq:33}) has no 
bound states. It follows that in the limit of small $m$, the EDM 
equations have no particlelike solutions, if $(e/m)^2 \geq 1$. This means 
that the mass-energy curves in Figure~\ref{fig2} can only start at 
$m=0$ if $(e/m)<1$. This is confirmed by the numerics 
(Figure~\ref{fig2}, curves C, D, and E). For $(e/m)^2 = 1$, the curve 
tends to $m-\omega=0$ as $m \rightarrow \infty$.

If $(e/m)^2 > 1$, Figure~\ref{fig2} shows that the EDM equations admit
solutions only if $m$ is sufficiently large, and smaller than some threshold
value where the binding energy of the fermions goes to zero.

\begin{figure}[tb]
        \epsfxsize=13cm
        \centerline{\epsfbox{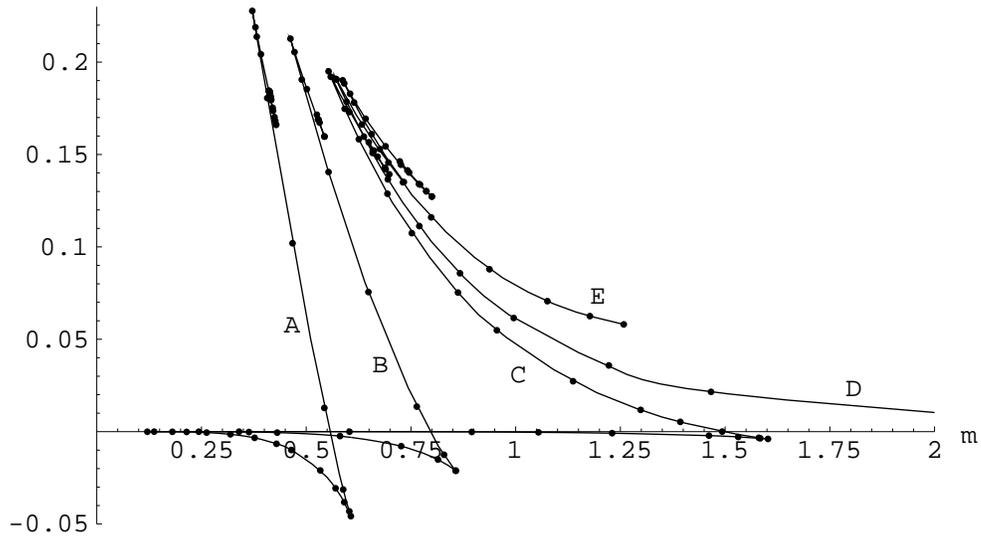}}
        \caption{Total Binding Energy $\rho-2m$ for $(e/m)^2=0$
        (A), $0.7162$ (B), $0.9748$ (C), $1$ (D), and $1.0313$ (E).}
        \label{fig3}
\end{figure}

We can also consider the total binding energy $\rho - 2m$, where 
$\rho$ is defined in (\ref{eq:23}). In Figure~\ref{fig3},
we plot $\rho-2m$ versus $m$, for various values of $(e/m)^2$. If 
$(e/m)^2<1$, $\rho-2m$ is negative for the stable solutions, while 
$\rho-2m>0$ if $(e/m)^2>1$. This indicates that if $(e/m)^2>1$, such 
solutions should be unstable because energy is gained by breaking up 
the binding.

\section{Non-Existence of Black Hole Solutions}
\label{sec5}
As we have noted in the last section, particlelike 
solutions of the EDM equations in a given state (e.g. the ground 
state) cease to exist if the rest mass $m$ of the fermions exceeds a 
certain threshold value $m_s$. The most natural physical 
interpretation of this statement is that if $m>m_s$, the 
gravitational interaction becomes so strong that a black hole would 
form. This suggests that there should be black hole solutions of the 
EDM equations for large fermion masses. In this section, we shall show 
that this intuitive picture of black hole formation is incorrect. In 
fact, we prove that under weak regularity conditions on the form of 
the horizon, any black hole solution of the EDM equations must either 
be the RN solution (in which case the Dirac wave function is 
identically zero), or the event horizon has the same general form as 
the extreme RN metric. In the latter case, we show numerically that 
the Dirac wave functions cannot be normalized. It follows that the EDM 
system does {\em{not}} admit black hole solutions. Thus the study of 
black holes in the presence of Dirac spinors leads to unexpected 
physical effects. If we apply this result to the gravitational 
collapse of a ``cloud'' of Dirac particles, our result indicates that 
the Dirac particles must eventually disappear inside the event horizon.

In order to establish these results, we first recall what is meant by
black hole solutions of the EDM equations. These are solutions of 
Equations (\ref{eq:E1})--(\ref{eq:E5}) defined in the region 
$r>\bar{r}>0$, which are asymptotically flat (so that (\ref{eq:22}) 
holds), and have finite (ADM) mass (so that (\ref{eq:23}) holds), and 
satisfy the normalization condition (\ref{eq:213}). In addition,
we assume that $A(r)>0$ for $r>\bar{r}$, and $\lim_{r \searrow \bar{r}}
A(r) = 0$, while $T(r)>0$ and $\lim_{r \searrow \bar{r}} T(r) = \infty$.

We make the following three assumptions on the regularity of the 
functions $A$, $T$, and $\phi$ on the form of the event horizon $r=\bar{r}$:
\begin{description}
\item[{\rm{(I)}}] The volume element $\sqrt{| {\mbox{det }} g_{ij}|} = r^2 
A^{-\frac{1}{2}}\: T^{-1}$ is smooth and non-zero on the horizon; 
i.e.,
\[ T^{-2} \:A^{-1},\: T^2 A \;\in\; C^\infty([\bar{r}, \infty)) \spc . \]
\item[{\rm{(II)}}] The electromagnetic field tensor is 
$F_{ij}=\partial_i A_j - \partial_j A_i$; we assume that the strength 
of the em field tensor $F_{ij} \:F^{ij} = -2 |\phi^\prime|^2 \:A\:T^2$
is bounded near the horizon. In view of (I), this means that we assume
\[ |\phi^\prime(r)| \;<\; c_1 \;\;\;,\spc \bar{r}<r<\bar{r}+\varepsilon \]
for some positive constants $c_1$, $\varepsilon>0$.
\item[{\rm{(III)}}] The function $A(r)$ obeys a power law; i.e.
\begin{equation}
        A(r) \;=\; c \:(r-\bar{r})^s \:+\: {\cal{O}}((r-\bar{r})^{s+1}) 
        \;\;\;,\spc r>\bar{r} \label{eq:41}
\end{equation}
for some positive constants $c$ and $s$.
\end{description}
A brief discussion of these assumptions is in order. Thus, if (I) or 
(II) were violated, then an observer freely falling into a black hole 
would feel strong forces when crossing the horizon. Assumption (III) 
is a technical condition which seems sufficiently general to include 
all physically relevant horizons; for example $s=1$ corresponds to the 
Schwarzschild horizon, and $s=2$ corresponds to the extreme RN 
horizon. However, (III) does not seem to be essential for our 
non-existence results, and with more mathematical effort, we believe 
that it could be weakened or even omitted completely.

Here is the main result in this section.
\begin{Thm}
\label{thm1}
Any black hole solution of the EDM equations (\ref{eq:E1})--(\ref{eq:E5})
which satisfies the regularity conditions (I)--(III) either is a 
non-extreme RN solution with $\alpha(r) \equiv 0 \equiv \beta(r)$, or 
$s=2$ and the following expansions are valid near the event horizon 
$r=\bar{r}$:
\begin{eqnarray}
A(r) &=& A_0 \:(r-\bar{r})^2 \:+\: {\cal{O}}((r-\bar{r})^3) \label{eq:p1} \\
T(r) &=& T_0 \:(r-\bar{r})^{-1} \:+\: {\cal{O}}((r-\bar{r})^0) 
\label{eq:p2} \\
\phi(r) &=& \frac{\omega}{e} \:+\: \phi_0 \:(r-\bar{r})
\:+\: {\cal{O}}((r-\bar{r})^2) \label{eq:p3} \\
\alpha(r) &=& \alpha_0 \:(r-\bar{r})^\kappa \:+\:
{\cal{O}}((r-\bar{r})^{\kappa+1}) \label{eq:p4a} \\
\beta(r) &=& \beta_0 \:(r-\bar{r})^\kappa \:+\:
{\cal{O}}((r-\bar{r})^{\kappa+1}) \label{eq:p4} \;\;\;,
\end{eqnarray}
with positive constants $A_0$, $T_0$, and real parameters 
$\phi_0$, $\alpha_0$, and $\beta_0$. The exponent $\kappa$ satisfies the 
constraint
\begin{equation}
        \frac{1}{2} \;<\; \kappa \;=\; A_0^{-1} \:\sqrt{m^2 - e^2 
        \phi_0^2 T_0^2 + \left( \frac{2j+1}{2 \bar{r}} \right)^2} \;\;\; ,
        \label{eq:p5}
\end{equation}
and the spinor coefficients $\alpha_0$ and $\beta_0$ are related by
\begin{equation}
        \alpha_0 \left( \sqrt{A_0} \:\kappa \:\pm\: \frac{2j+1}{2 \bar{r}} 
        \right) \;=\; -\beta_0 \:(m - e \phi_0 T_0) \;\;\; ,
        \label{eq:p6}
\end{equation}
where `$\pm$' refers to the two choices of the signs in
(\ref{eq:E1})--(\ref{eq:E5}).
\end{Thm}

We shall now outline a proof of this result; we first consider the 
case that the exponent $s<2$ in (\ref{eq:41}).
\begin{Lemma}
\label{lemma1}
Assume that $s<2$ and that $(\alpha, \beta, A, T, \phi)$ is a black-hole 
solution where $(\alpha, \beta) \not \equiv 0$. Then there are 
constants $c, \varepsilon>0$ satisfying
\begin{equation}
        c \;\leq\; \alpha(r)^2 + \beta(r)^2 \;\leq\; \frac{1}{c} 
        \;\;\;,\spc \bar{r} < r < \bar{r}+\varepsilon \;\;\; .
        \label{eq:n7}
\end{equation}
\end{Lemma}
{\Proof}
According to (\ref{eq:E1}) and (\ref{eq:E2}), we have
\begin{eqnarray}
\sqrt{A} \:\frac{d}{dr} (\alpha^2 + \beta^2) &=& 2
\left( \begin{array}{c} \alpha \\ \beta \end{array} \right)
\left( \begin{array}{cc} \displaystyle \pm\frac{2j+1}{2r} & -m \\
-m & \displaystyle \mp\frac{2j+1}{2r} \end{array} \right)
\left( \begin{array}{c} \alpha \\ \beta \end{array} \right) \nonumber \\
&\leq& \left(4 m^2 + \frac{(2j+1)^2}{r^2} \right)^{\frac{1}{2}} (\alpha^2 + 
\beta^2) \;\;\; . \label{eq:n3}
\end{eqnarray}
The uniqueness theorem for ODEs implies that $(\alpha^2 + 
\beta^2)(r)>0$ for all $r$, $\bar{r} < r < \bar{r}+\varepsilon$, for 
any $\varepsilon>0$. Dividing (\ref{eq:n3}) by $\sqrt{A} 
\:(\alpha^2 + \beta^2)$ and integrating from $r>\bar{r}$ to 
$\bar{r}+\varepsilon$ gives
\begin{eqnarray}
\lefteqn{ \left| \log ((\alpha^2 + \beta^2)(\bar{r} + \varepsilon)) \:-\:
\log ((\alpha^2 + \beta^2)(r)) \right| } \nonumber \\
&\leq& \int_r^{\bar{r} + \varepsilon}
A^{-\frac{1}{2}}(t) \left(4 m^2 + \frac{(2j+1)^2}{t^2} 
\right)^{\frac{1}{2}} dt \;\;\;.
        \label{eq:n4}
\end{eqnarray}
Since $s<2$, (\ref{eq:41}) implies that $A^{-\frac{1}{2}}$ is 
integrable on $\bar{r} \leq r \leq \bar{r} + \varepsilon$, so that
the integral in (\ref{eq:n4}) is majorized by
\[ \int_{\bar{r}}^{\bar{r} + \varepsilon}
A^{-\frac{1}{2}}(t) \left(4 m^2 + \frac{(2j+1)^2}{t^2} 
\right)^{\frac{1}{2}} dt \;\;\;, \]
and this yields (\ref{eq:n7}).
\QED

We can now dispose of the case $0<s<2$; namely, we have
\begin{Prp}
\label{prp1}
If $0<s<2$, then the only black hole solutions of the system 
(\ref{eq:E1})--(\ref{eq:E5}) are the non-extreme Reissner-Nordstr\"om 
solutions.
\end{Prp}
{\Proof}
We assume that we have a solution such that $(\alpha, \beta)(r) \not 
\equiv 0$, and show that this gives a contradiction.

The last lemma implies that the spinors are bounded near $r=\bar{r}$. 
From (\ref{eq:E3}) and (\ref{eq:E4}), we find
\begin{eqnarray}
r \:\frac{d}{dr}(A T^2) &=& -4(2j+1) \:(\omega - e \phi)
\:T^4\:(\alpha^2+\beta^2) \:\pm\: 2 \:\frac{(2j+1)^2}{r} 
\:T^3\:\alpha \beta \nonumber \\
&&+2 (2j+1) \:m \:T^3\: (\alpha^2 - \beta^2) \;\;\; .
\label{eq:nca}
\end{eqnarray}
Assumption (I) implies that the left side of (\ref{eq:nca}) is regular 
so the same is true of the right side. Since $T \rightarrow \infty$ 
as $r \searrow \bar{r}$, we conclude that
\begin{equation}
\lim_{\bar{r} < r \rightarrow \bar{r}} (\omega - e \phi(r)) \;=\; 0 \;\;\; .
        \label{eq:n8}
\end{equation}
From Maxwell's equation
\begin{equation}
\phi^{\prime \prime} \;=\; -\frac{1}{A} \:\frac{(2j+1)\: e}{r^2}
\:(\alpha^2+\beta^2) \:-\: \frac{1}{r^2 \:\sqrt{A} \:T} \:[r^2 
\:\sqrt{A} \:T]^\prime \:\phi^\prime \;\;\;,
        \label{eq:n5}
\end{equation}
we see that (I) implies that the coefficient of $\phi^\prime$ is smooth. 
If $s \geq 1$, $A^{-1}$ is not integrable at $\bar{r}$, so that 
$|\phi^\prime|$ is unbounded at $\bar{r}$, thereby contradicting (II). 
Thus $s<1$, and integrating (\ref{eq:n5}) twice and using 
(\ref{eq:n8}) gives near $r=\bar{r}$ the following expansions:
\begin{eqnarray*}
\phi^\prime(r) &=& c_1 \:(r-\bar{r})^{-s+1} \:+\: c_2 \:+\: 
{\cal{O}}((r-\bar{r})^{-s+2}) \;\;\;{\mbox{, and}} \\
\phi(r) &=& c_1 \:(r-\bar{r})^{-s+2} \:+\: c_2 \:(r-\bar{r}) \:+\: 
\frac{\omega}{e} \:+\: {\cal{O}}((r-\bar{r})^{-s+3}) \;\;\;.
\end{eqnarray*}
Using these in (\ref{eq:E3}), and noting that $A$ and $r^2 A T^2 
\:|\phi^\prime|^2$ are bounded near $r=\bar{r}$, and that
$(\omega - e \phi) = {\cal{O}}(r-\bar{r})$, and $T^2 
(\alpha^2 + \beta^2) \sim (r-\bar{r})^{-s}$, $s<1$, we see that the rhs
of (\ref{eq:E3}) is bounded near $r=\bar{r}$. On the other 
hand, the lhs of (\ref{eq:E3}) diverges near $r=\bar{r}$ since $r 
A^\prime(r) = (r-\bar{r})^{-s+1}$; this contradiction completes the proof.
\QED
In the case $s \geq 2$, we first prove the following two facts 
(cf.~\cite{FSY3}):
\begin{equation}
\lim_{r \searrow \bar{r}} (r-\bar{r})^{-\frac{s}{2}} \:(\alpha^2 + 
\beta^2) \;=\; 0 \label{eq:413}
\end{equation}
and
\begin{equation}
\lim_{r \searrow \bar{r}} |\phi^\prime(r)| \;=\; \bar{r}^{-1}\: \lim_{r 
\searrow \bar{r}} A^{-\frac{1}{2}} \:T^{-1} \;>\; 0 \;\;\;.
\label{eq:414}
\end{equation}
From (\ref{eq:414}), we find that
\[ (w - e \phi)(r) \;=\; c \:+\: d \:(r-\bar{r}) \:+\: o(r-\bar{r})\;\;\;, \]
where $d = e/\bar{r} \:\lim_{r \searrow \bar{r}} A^{-\frac{1}{2}} 
\:T^{-1} > 0$. Thus $(\omega - e \phi)T$ diverges monotonically. 
From (\ref{eq:E1}) and (\ref{eq:E2}), this implies that $\liminf_{r 
\searrow \bar{r}} (\alpha^2 + \beta^2) > 0$, thereby 
contradicting (\ref{eq:413}). Thus if $s>2$, there are no solutions 
of (\ref{eq:E1})--(\ref{eq:E5}).\\[.5em]
{\em{Proof of Theorem \ref{thm1}: }}
We must only consider the case that $s=2$ and (\ref{eq:p1}), 
(\ref{eq:p2}) hold. From (\ref{eq:413}) we see that $\lim_{r \searrow 
\bar{r}} \alpha^2 + \beta^2 = 0$, and we can show that $(\omega - e 
\phi)T$ cannot diverge monotonically near $r=\bar{r}$ (see \cite{FSY4}). 
But (\ref{eq:414}) shows that $(\omega - e \phi)$ has a Taylor 
expansion near $r=\bar{r}$ with a non-zero linear term. Thus 
(\ref{eq:414}) holds, the constant term in the Taylor expansion of 
$(\omega - e \phi)$ vanishes, and $\lim_{r \searrow \bar{r}}(\omega - e 
\phi) T = \lambda$, where from (\ref{eq:414}), $|\lambda|=\bar{r}^{-1} 
\:\lim_{r \searrow \bar{r}} A^{-\frac{1}{2}} \:T^{-1} > 0$. As 
in~\cite{FSY4}, we may write the Dirac equations in the variable
\[ u(r) \;=\; -r-\bar{r} \:\ln(r-\bar{r}) \]
and apply the stable manifold theorem to conclude that 
$\alpha$ and $\beta$ satisfy the power laws (\ref{eq:p4a}),(\ref{eq:p4}), 
and (\ref{eq:413}) yields that $K>\frac{1}{2}$. Using 
(\ref{eq:p1})--(\ref{eq:p4}) into (\ref{eq:E1}) and (\ref{eq:E2}) 
gives
\begin{eqnarray*}
\sqrt{A_0} \:\kappa\:\alpha_0 &=& \pm\frac{2j+1}{2 \bar{r}} \:\alpha_0 
\:+\: (e \phi_0 T_0 - m) \:\beta_0 \\
\sqrt{A_0} \:\kappa\:\beta_0 &=& -(e \phi_0 T_0 + m) \alpha_0
\:\mp\: \frac{2j+1}{2 \bar{r}} \:\beta_0 \;\;\; ,
\end{eqnarray*}
which are equivalent to (\ref{eq:p5}) and (\ref{eq:p6}). This 
completes the proof of Theorem~\ref{thm1}.
\QED

Notice that in the case of non-zero spinors ($s=2$), 
Theorem~\ref{thm1} places severe constraints on the behavior of black 
hole solutions near the event horizon, in the sense that since 
$\kappa > \frac{1}{2}$, the spinors decay so fast at $r=\bar{r}$, that 
both the metric and the em field behave like the extreme RN solution 
on the event horizon. Physically speaking, this restriction to the 
extremal case means that the electric charge of the black hole is so 
large that the electric repulsion balances the gravitational 
attraction, and prevents the Dirac particles from ``falling into'' the 
black hole. Of course, this is not the physical situation that one 
expects in the gravitational collapse of, say, a star. However, 
extreme RN black holes are physically important since they have zero 
temperature \cite{H}, and can be considered to be the asymptotic 
states of black holes emitting Hawking radiation. It is thus 
interesting to see if the expansions (\ref{eq:p1})--(\ref{eq:p4}) 
yield global black hole solutions of the EDM equations.

This question is especially interesting since in the next section we 
shall show that for an extreme RN background field, spinors 
satisfying the expansions (\ref{eq:p4a}), (\ref{eq:p4}) cannot be 
normalized. The question thus becomes whether the influence of the 
spinors on the gravitational and em field can yield black hole 
solutions with normalized spinors. This is a very difficult question 
because one must analyze the global behavior of these solutions of 
the EDM equations. Our numerical investigations show that the answer 
to the above question is negative; namely solutions either develop a 
singularity for some $r>\bar{r}$, or the spinors $(\alpha, \beta)$ are 
not normalizable. We thus conclude that the expansions
(\ref{eq:p1})--(\ref{eq:p4}) do not give normalizable solutions of the
EDM equations.

\section{Dirac Particles in a Reissner-Nordstr\"om Background}
In this section, we shall consider solutions of the EDM equations 
where we fix the background metric and em field to be an RN solution. 
Near a collapsing black hole one might guess that Dirac particles 
might get into a static or time periodic state. However, we shall 
show that, in contrast to the classical situation, the Dirac equations do
not admit any normalizable time-periodic 
solutions; in particular, they admit no 
normalizable static solutions. We do not assume any spatial symmetry 
on the wave functions. This result can be physically interpreted as 
saying that Dirac particles can either disappear into the black hole 
of escape to infinity, but they cannot remain on a periodic orbit 
around the black hole. We note that it is essential for our arguments 
that the particles have spin. In fact, in the case where the particles 
do not have spin, the Dirac equation must be replaced by the 
Klein-Gordon equation, and our arguments fail; c.f.~\cite{L}.

The RN metric can be written in polar coordinates as
\begin{equation}
ds^2 \;=\;  \left(1 - \frac{2 \rho}{r} \:+\: 
        \frac{q^2}{r^2}\right) dt^2 \:-\: \left(1 - \frac{2 \rho}{r} \:+\: 
        \frac{q^2}{r^2}\right)^{-1} dr^2 \:-\: r^2 
        \:(d\vartheta^2 + \sin^2 \vartheta \:d\varphi^2) \;\;\; ,
        \label{eq:51}
\end{equation}
where $\rho$ is the (ADM) mass of the black hole, and $q$ its charge. 
The em potential is of the form $(-\phi, \vec{0})$ with Coulomb 
potential
\begin{equation}
\phi(r) \;=\; \frac{q}{r} \;\;\;.
        \label{eq:52}
\end{equation}
In the ``non-extremal'' case ($q<\rho$), the metric coefficient 
$(1-\frac{2 \rho}{r} + \frac{q^2}{r^2})$ vanishes twice, and thus 
there are two horizons $0<r_0<r$. If $q=\rho$, the metric is called an extreme 
Reissner-Nordstr\"om (ERN) metric and has a single horizon at 
$r=\rho$. If $q>\rho$, the above metric coefficient is non-vanishing, 
and so the metric does not describe a black hole; this case will not 
be considered.

We consider time-periodic solutions, noting that static solutions are 
a special case. Since the phase of the Dirac wave function $\Psi$ has 
no physical significance, we define $\Psi$ to be periodic with period 
$T$ if for some real $\Omega$,
\begin{equation}
\Psi(t+T, r, \vartheta, \varphi) \;=\; e^{-i \Omega T} 
\:\Psi(t,r,\vartheta, \varphi) \;\;\;.
        \label{eq:53}
\end{equation}
Our main theorem in this section is the following:
\begin{Thm}
\label{thm2}
{\em{i)}} In a non-extreme RN background, there are no normalizable, 
time-periodic solutions of the Dirac equation. {\em{ii)}} In an ERN 
background, every normalizable, time-periodic solution of the Dirac 
equation is identically zero in the region $r>\rho$.
\end{Thm}

We shall begin by deriving conditions which relate the wave function 
$\Psi$ on both sides of the event horizon. We first consider the 
case of a non-extreme RN background, and analyze the behavior of 
$\Psi$ near the event horizon. For this, we begin by studying the 
behavior of $\Psi$ in a Schwarzschild background metric, and we shall 
also consider the Dirac equation in different coordinate systems. This 
is done with the aim of passing to Kruskal coordinates, in order to 
remove the ``Schwarzschild singularity.''

The Schwarzschild metric is
\[ ds^2 \;=\;   \left(1 - \frac{2 \rho}{r} \right) dt^2 \:-\:
\left(1 - \frac{2 \rho}{r} \right)^{-1} dr^2 \:-\: r^2 
        \:(d\vartheta^2 + \sin^2 \vartheta \:d\varphi^2) \;\;\; , \]
where $\rho$ is the (ADM) mass, and the event horizon is at $r=2 \rho$. 
Some straightforward calculations (see~\cite{FSY4}) shows that outside 
the horizon ($r>\rho$), the Dirac operator can be written as
\begin{equation}
        G_{\mbox{\scriptsize{out}}}
        \;=\; \frac{i}{S} \:\gamma^t \:\frac{\partial}{\partial t} \:+\: 
        \gamma^r \left(i S\: \frac{\partial}{\partial r} \:+\: 
        \frac{i}{r} \:(S-1) \:+\: \frac{i}{2} \:S^\prime \right) 
        \:+\: i \gamma^\vartheta \frac{\partial}{\partial \vartheta}
        \:+\: i \gamma^\varphi \frac{\partial}{\partial \varphi} \;\;\; ,
        \label{eq:54}
\end{equation}
where
\[ S(r) \;=\; \left| 1 - \frac{2 \rho}{r} \right|^{\frac{1}{2}} 
\;\;\;. \]
The normalization integral is considered over the hypersurface 
$t={\mbox{const}}$; i.e.
\begin{equation}
(\Psi \:|\: \Psi)^t_{\mbox{\scriptsize{out}}} \;:=\;
\int_{\sR^3 \setminus B_{2 \rho}} 
        (\overline{\Psi} \gamma^t \Psi)(t, \vec{x}) \:S^{-1}\: d^3 x 
        \;\;\; , \label{eq:55}
\end{equation}
where $B_{2 \rho}$ denotes the ball of radius $2 \rho$ about the 
origin, and $\overline{\Psi}=\Psi^* \gamma^0$ is the adjoint spinor. 
In the region $r<2 \rho$, the Dirac operator is given by
\[ G_{\mbox{\scriptsize{in}}}
        \;=\; \gamma^r \left( \frac{i}{S} \:\frac{\partial}{\partial t} 
        \:-\: \frac{i}{r} \right) \:-\: \gamma^t \left( i S 
        \:\frac{\partial}{\partial r} \:+\: \frac{i}{r} \:S \:+\: 
        \frac{i}{2} \: S^\prime \right)
        \:+\: i \gamma^\vartheta \frac{\partial}{\partial \vartheta}
        \:+\: i \gamma^\varphi \frac{\partial}{\partial \varphi} \]
with corresponding normalization integral
\begin{equation}
        ( \Psi \:|\: \Psi)_{\mbox{\scriptsize{in}}}^t \;:=\;
        \int_{B_{2\rho}} (\overline{\Psi} \gamma^r \Psi)(t, \vec{x}) \;S^{-1}\;
        d^3 x \;\;\;. \label{eq:56}
\end{equation}

\begin{figure}[tb]
        \centerline{\epsfbox{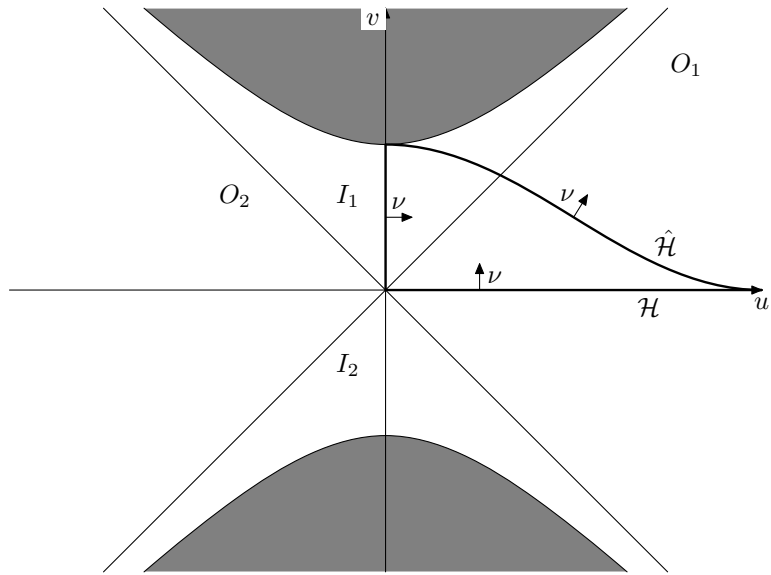}}
        \caption{Kruskal Coordinates}
        \label{fig4}
\end{figure}

Our description of spinors in this coordinate system poses certain 
difficulties. Namely, since the $t$ variable is space-like inside the 
horizon, the normalization integral (\ref{eq:56}) is not definite 
since the integrand is not positive. Thus we can no longer interpret 
the integrand as a probability density. Moreover, the Dirac equations 
corresponding to the operators $G_{\mbox{\scriptsize{in}}}$ and
$G_{\mbox{\scriptsize{out}}}$ describe the wave functions inside and 
outside the horizon, respectively. But it is not evident how to match 
the wave functions on the horizon.
To handle these issues, we remove the singularity at $r=2 \rho$ by 
going over to Kruskal coordinates.
Recall (see~\cite{ABS}) that Kruskal coordinates $u$ and $v$ are 
defined by
\begin{eqnarray}
u & = & \left\{ \begin{array}{ll}
\displaystyle \sqrt{\frac{r}{2\rho}-1} \:e^{\frac{r}{4\rho}} \:\cosh \left(
\frac{t}{4\rho}\right) & {\mbox{for $r>2\rho$}} \\[.8em]
\displaystyle \sqrt{1-\frac{r}{2\rho}} \:e^{\frac{r}{4\rho}} \:\sinh \left(
\frac{t}{4\rho}\right) & {\mbox{for $r<2\rho$}} \end{array} \right. 
\label{eq:57} \\[.3em]
v & = & \left\{ \begin{array}{ll}
\displaystyle \sqrt{\frac{r}{2\rho}-1} \:e^{\frac{r}{4\rho}} \:\sinh \left(
\frac{t}{4\rho}\right) & {\mbox{for $r>2\rho$}} \\[.8em]
\displaystyle \sqrt{1-\frac{r}{2\rho}} \:e^{\frac{r}{4\rho}} \:\cosh \left(
\frac{t}{4\rho}\right) & {\mbox{for $r<2\rho$}}  \spc . \end{array} \right.
\label{eq:58}
\end{eqnarray}
The horizon $r=2 \rho$ maps to the origin $u=0=v$, and the 
singularity $r=0$ maps to the hyperbola $v^2-u^2=1$, $v>0$. In Kruskal 
coordinates, the metric (\ref{eq:51}) becomes
\[ ds^2 \;=\; f^{-2} \:(dv^2 - du^2) \:-\: r^2(d\vartheta^2 + \sin^2 
\vartheta \: d\varphi^2) \]
where $f^{-2} \;=\; \frac{32 \rho^3}{r} \: e^{-\frac{r}{2\rho}}$.
Taking $v$ and $u$ as time and space variables, respectively, and 
noting that the metric is regular at the origin, we can extend the 
Dirac operator smoothly across the origin. A straightforward 
computation gives the Dirac operator in Kruskal coordinates as
\begin{eqnarray}
G &=& \gamma^t \left( f i\:\frac{\partial}{\partial v} \:+\: \frac{i}{r} \:
f \:(\partial_v r)\:-\: \frac{i}{2} \:\partial_v f \right)
\:+\: \gamma^r \left( fi\:\frac{\partial}{\partial u} \:+\: \frac{i}{r}
\:(f \:(\partial_u r)-1) \:-\:\frac{i}{2} \:\partial_u f \right) \nonumber \\
&&+\: i \gamma^\vartheta \partial_\vartheta
                + i \gamma^\varphi \partial_\varphi \;\;\; .
        \label{eq:59}
\end{eqnarray}
Observe that the Dirac operator is smooth across the event horizon.
Moreover, the normalization integrals (\ref{eq:55}) and (\ref{eq:56}) on the 
surface $t=0$ become
\[ (\Psi \:|\: \Phi) \;=\; \int_{\cal{H}} \overline{\Psi} G^j \Phi 
        \:\nu_j \:d\mu \;\;\;, \]
where
\[ {\cal{H}} \;=\; \{u=0,\:0 \leq v \leq 1\} \cup \{v=0,\:u>0 \} \;\;\;, \]
$\nu$ is the normal to $H$ pointing into the region $u>0$, $v>0$, 
and $G^j$ are the Dirac matrices
\[ G^v = f \:\gamma^t \;\;,\;\;\;\; G^u = f
\:\gamma^r \;\;,\;\;\;\; G^\vartheta = \gamma^\vartheta 
\;\;,\;\;\;\; G^\varphi = \gamma^\varphi \;\;\; . \]
We remark that for smooth solutions of the Dirac equation, one can use
current conservation
\begin{equation}
\nabla_j \overline{\Psi} G^j \Psi \;=\; 0 \;\;\;,
        \label{eq:510}
\end{equation}
to continuously deform the hypersurface ${\cal{H}}$ keeping fixed 
the value of the normalization integral. For example, one can
deform ${\cal{H}}$ to $\hat{\cal{H}}$ as depicted in Figure~\ref{fig4},
thereby avoiding 
integrating across the horizon. On the other hand, 
one must exercise extreme care whenever a solution of the Dirac equation is 
singular near the origin.

As shown in~\cite{FSY4}, the Dirac operator in Kruskal coordinates 
can be written as
\begin{equation}
        G \;=\; U \:G_{\mbox{\scriptsize{out}}}\:U^{-1} \;=\;
        U \:G_{\mbox{\scriptsize{in}}}\:U^{-1} \;\;\; ,
        \label{eq:511}
\end{equation}
where $U$ is the time-dependent matrix
\begin{equation}
U(t) \;=\; \cosh \left(\frac{t}{8 \rho}\right) \:\1 \:+\: \sinh 
\left(\frac{t}{8 \rho}\right) \:\gamma^t \:\gamma^r \;\;\; ,
        \label{eq:512}
\end{equation}
and the Dirac operators $G_{\mbox{\scriptsize{out}}}$ and
$G_{\mbox{\scriptsize{in}}}$ in Kruskal coordinates are
\begin{eqnarray*}
G_{\mbox{\scriptsize{out}}} &=& \frac{i}{4 \rho S} \:(u 
\gamma^t + v \gamma^r) \:\frac{\partial}{\partial v} \:+\: 
\frac{i}{4 \rho S} \:(v \gamma^t + u \gamma^r) \:\frac{\partial}{\partial u} \\
&&+ \left( \frac{i}{r}\:(S-1) \:+\:\frac{i}{2} \:S^\prime 
\right) \gamma^r \:+\: i \gamma^\vartheta \:\frac{\partial}{\partial 
\vartheta} \:+\: i \gamma^\varphi \:\frac{\partial}{\partial 
\varphi} \\
G_{\mbox{\scriptsize{in}}} &=& \frac{i}{4\rho S} \:(v 
\gamma^t + u \gamma^r) \:\frac{\partial}{\partial v} \:+\: 
\frac{i}{4 \rho S} \:(u \gamma^t + v \gamma^r) \:\frac{\partial}{\partial u} \\
&&- \left( \frac{i}{r}\:S \:+\:\frac{i}{2} \:S^\prime 
\right) \gamma^t \:-\:\frac{i}{r} \:\gamma^r\:+\: i \gamma^\vartheta \:\frac{\partial}{\partial 
\vartheta} \:+\: i \gamma^\varphi \:\frac{\partial}{\partial \varphi} \;\;\;.
\end{eqnarray*}
It follows that the Dirac operators $G_{\mbox{\scriptsize{out}}}$ and
$G_{\mbox{\scriptsize{in}}}$ can be identified with the Dirac 
operator $G$ in the region
\[ {\cal{R}} \;=\; \{ u+v>0,\; v^2-u^2 < 1 \} \;\;\;. \]

We next see how solutions of the Dirac equation inside and outside the 
horizon match on the horizon, $u=0=v$. To do this, we first study the 
behavior of these solutions on the horizon. Let us first consider 
{\em{static}} solutions of the Dirac equation, so
\[ \Psi(t,r,\vartheta,\varphi) \;=\; e^{-i \omega t} 
\:\Psi(r,\vartheta,\varphi) \;\;\;. \]
We assume that $\Psi$ is a solution of the Dirac equations
$(G_{\mbox{\scriptsize{in}}}-m)=0$ and $(G_{\mbox{\scriptsize{out}}}-m)=0$,
and that $\Psi$ is smooth on both sides of the horizon $r<2 \rho$ 
and $r>2 \rho$. Using (\ref{eq:511}) and (\ref{eq:512}), we have
\[ \Psi(u,v,\vartheta,\varphi) \;=\; U(t) \:e^{-i \omega t} 
\:\Psi(r,\vartheta, \varphi) \;\;\;, \]
where $r$ and $t$ are determined implicitly from $u$ and $v$ in the 
usual way (see~\cite{ABS}). This implies that $\Psi$ is only defined 
in ${\cal{R}}$, and solves there the Dirac equation $(G-m) \Psi = 0$. 
Since we are only considering black holes, we demand that $\Psi$ 
vanishes in the half-plane $u+v<0$; thus we must analyze solutions 
$\Psi$ of the form
\[ \Psi(u,v,\vartheta, \varphi) \;=\; \left\{ \begin{array}{cl}
U(t) \: e^{-i \omega t} \:\Psi(r, \vartheta, \varphi)
& {\mbox{for $u+v>0$, $u \neq v$}} \\ 0
& {\mbox{for $u+v<0 \;\;\; .$}}
\end{array} \right. \]
Such a wave function might be singular along the lines $u=\pm v$, in 
which case $\Psi$ must satisfy the Dirac equation in a generalized 
sense. An analysis carried out in~\cite{FSY4} shows that $\Psi$ must 
satisfy the two matching conditions
\begin{eqnarray}
\lefteqn{ \lim_{\varepsilon \rightarrow 0} (\gamma^t + 
\gamma^r) \:|\varepsilon|^{\frac{1}{4}}
\Psi(t, 2\rho + \varepsilon, \vartheta, \varphi) \;=\; 0 }
\label{eq:513} \\
\lefteqn{ |\varepsilon|^{\frac{1}{4}} \left(
\Psi(t,2 \rho + \varepsilon, \vartheta, \varphi) \:-\: \Psi(t,2 \rho -
\varepsilon, \vartheta, \varphi) \right) } \nonumber \\
&=& o(1+|\varepsilon|^{\frac{1}{4}} \:\Psi(t,2 \rho + \varepsilon, \vartheta,
\varphi)) \;\;\;{\mbox{as $\varepsilon \rightarrow 0$.}} \label{eq:514}
\end{eqnarray}
Note that since these only depend on the local behavior of $\Psi$ near 
the horizon, they are also applicable when we are in the case of a 
non-extreme RN background having event horizon at $r=2 \rho$.

We now consider Dirac particles in a RN background. Since the 
gravitational and EM background
fields are spherically symmetric and time independent, we 
can separate out the angular and time dependence of the wave functions 
via spherical harmonics and plane waves in the usual manner and, as 
shown in \cite{FSY4}, we obtain the following two component Dirac 
equations: In regions where the $t$-variable is time-like,
\begin{eqnarray}
\lefteqn{ S \:\frac{d}{dr} \Phi^\pm_{jk \omega} } \nonumber \\
&=& \left[ \left( 
\begin{array}{cc} 0 & -1 \\ 1 & 0 \end{array} \right) (\omega - e 
\phi) \:\frac{1}{S} \:\pm\:\left( \begin{array}{cc} 1 & 0 \\ 0 & -1 \end{array}
\right) \frac{2j+1}{2r} \:-\: \left(
\begin{array}{cc} 0 & 1 \\ 1 & 0 \end{array} \right) m \right]
\Phi^\pm_{jk \omega} \;\;\; , \spc
        \label{eq:516}
\end{eqnarray}
and in the regions where $t$ is space-like,
\begin{eqnarray}
\lefteqn{ S \:\frac{d}{dr} \Phi^\pm_{jk \omega} } \nonumber \\
&=& \left[ \left( 
\begin{array}{cc} 0 & -1 \\ 1 & 0 \end{array} \right) (\omega - e 
\phi) \:\frac{1}{S} \:\pm\:i \left( \begin{array}{cc} 0 & 1 \\ 1 & 0
\end{array} \right) \frac{2j+1}{2r} \:+\:i \left(
\begin{array}{cc} 1 & 0 \\ 0 & -1 \end{array} \right) m \right]
\Phi^\pm_{jk \omega} \;\;\; . \spc
        \label{eq:517}
\end{eqnarray}
In these equations,
\begin{equation}
S(r) \;=\; \left| 1 \:-\: \frac{2 \rho}{r} \:+\: \frac{q^2}{r^2} 
\right|^{\frac{1}{2}} \;\;\;
        \label{eq:518}
\end{equation}
$j=\frac{1}{2}, \frac{3}{2},\ldots$, $k=-j,-j+1,\ldots,j$, and the 
$\pm$ signs correspond as before to the two eigenvalues of the operator
$\gamma^0 P$ (cf.\ Section~\ref{sec2}). Here we have chosen for the 
Dirac wave functions the two ansatz'
\begin{eqnarray}
\Psi^+_{jk \:\omega} &=& e^{-i \omega t} \:\frac{S^{-\frac{1}{2}}}{r}\:
        \left( \begin{array}{c} \chi^k_{j-\frac{1}{2}} \:\Phi^+_{jk \omega \:1}(r) \\
i \chi^k_{j+\frac{1}{2}}\:\Phi^+_{jk \omega \:2}(r) \end{array} \right)
\label{eq:519} \\
\Psi^-_{jk \:\omega} &=& e^{-i \omega t} \:\frac{S^{-\frac{1}{2}}}{r}\:
        \left( \begin{array}{c} \chi^k_{j+\frac{1}{2}} \:\Phi^-_{jk \omega \:1}(r) \\
i \chi^k_{j-\frac{1}{2}}\:\Phi^-_{jk \omega \:2}(r) \end{array} 
\right) \label{eq:520}
\end{eqnarray}
with 2-spinors $\Phi^\pm_{jk \omega}$, and $\chi^k_{j \pm \frac{1}{2}}$ 
are defined by
\begin{eqnarray}
\chi^k_{j-\frac{1}{2}} &=& \sqrt{\frac{j+k}{2j}} 
\:Y^{k-\frac{1}{2}}_{j-\frac{1}{2}} \: \left( 
\begin{array}{c} 1 \\ 0 \end{array} \right) \:+\:
\sqrt{\frac{j-k}{2j}} \:Y^{k+\frac{1}{2}}_{j-\frac{1}{2}} \: \left( 
\begin{array}{c} 0 \\ 1 \end{array} \right) \label{eq:521} \\
\chi^k_{j+\frac{1}{2}} &=& \sqrt{\frac{j+1-k}{2j+2}} 
\:Y^{k-\frac{1}{2}}_{j+\frac{1}{2}} \: \left( 
\begin{array}{c} 1 \\ 0 \end{array} \right) \:-\:
\sqrt{\frac{j+1+k}{2j+2}} \:Y^{k+\frac{1}{2}}_{j+\frac{1}{2}} \: \left( 
\begin{array}{c} 0 \\ 1 \end{array} \right) \label{eq:522} \;\;\; .
\end{eqnarray}
where $Y^k_l$ are the usual spherical harmonics, $l=0,1,2,\ldots$, 
$k=-l,\ldots,l$.

We shall show that the matching conditions (\ref{eq:513}),(\ref{eq:514})
do not yield normalizable, time-periodic solutions of the Dirac equation. 
This will be done by showing that, for every non-zero solution of 
Dirac's equation, the normalization integral outside and away from 
the horizons
\begin{equation}
(\Psi \:|\: \Psi)^t \;=\; \int_{\sR^3 \setminus B_{2 r_1}} 
\overline{\Psi} \gamma^t \Psi \;S^{-1}\:d^3x \;\;\;,
        \label{eq:523}
\end{equation}
is infinite for some $t$. Note that for a normalizable wave function, 
this integral is the probability that the particle lies outside the 
ball of radius $r_1$, and thus cannot exceed $1$. So if 
(\ref{eq:523}) is infinite, the wave function cannot be normalized.

Now assume that $\Psi$ is a $T$-periodic solution of Dirac's 
equation. Expanding the periodic function $e^{-i \Omega t} 
\:\Psi(t,r,\vartheta, \varphi)$ in a Fourier series, and using the 
basis (\ref{eq:517}),(\ref{eq:518}) yields
\begin{equation}
\Psi(t,r,\vartheta,\varphi) \;=\; \sum_{n, j, k, s} \Psi^s_{j k 
\:\omega(n)}(t,r, \vartheta,\varphi) \;\;\; , \label{eq:524}
\end{equation}
where $s=\pm$, and $\omega(n) = \Omega + 2 \pi n/T$. Using the 
orthonormality of the spinors $\chi^k_{j \pm \frac{1}{2}}$, the 
integral (\ref{eq:521}) becomes
\[ (\Psi \:|\: \Psi)^t \;=\; \int_{\sR^3 \setminus B_{2 r_1}}
\sum_{n, n^\prime} \sum_{j,k,s} \overline{\Psi^s_{j k\:\omega(n)}}
\gamma^t \Psi^s_{j k\:\omega(n^\prime)} \:S^{-1}\:d^3x \;\;\;. \]
In order to eliminate the oscillating time dependence of the 
integrand, we average over one period $(0,T)$ to get
\[ \frac{1}{T} \int_0^T (\Psi \:|\: \Psi)^t \:dt \;=\; 
\sum_{n, j, k, s} \:(\Psi^s_{j k \:\omega(n)} \:|\: \Psi^s_{j k 
\:\omega(n)}) \;\;\;. \]
For a normalizable wave function, this expression is finite, and 
hence all summands are finite: For all $s$, $j$, $k$, $n$,
\begin{equation}
( \Psi^s_{j k \:\omega(n)} \:|\: \Psi^s_{j k \:\omega(n)}) \;<\; 
\infty \;\;\;. \label{eq:525}
\end{equation}
We shall show that (\ref{eq:525}) cannot hold for non-trivial 
solutions of the Dirac equation; for this we begin with
\begin{Lemma}
\label{lemma5}
The function $|\Phi^\pm_{jk\omega}(r)|^2$ has finite boundary values on 
both horizons, and if it is zero on one horizon, then it is identically 
zero.
\end{Lemma}
{\Proof}
For simplicity, we omit the indices $j$, $k$, and $\omega$.
Choose $\delta$, $0<\delta<r_0$, and notice that the $t$-direction is 
time-like on the intervals $(\delta, r_0)$ and $(r_1, \infty)$. In 
these regions, we can use (\ref{eq:516}) to obtain
\begin{eqnarray*}
S \:\frac{d}{dr} |\Phi^\pm|^2(r) &=& \bra S \:\frac{d}{dr} \Phi^\pm,\: \Phi^\pm \ket
\:+\: \bra \Phi^\pm,\: S \:\frac{d}{dr} \Phi^\pm \ket \\
&=& \pm\frac{2j+1}{r} 
\:(|\Phi^\pm_1|^2 - |\Phi^\pm_2|^2) \:-\: 4 m
\:{\mbox{Re}}\left((\Phi^\pm_1)^* \:\Phi^\pm_2 \right) \;\;\;,
\end{eqnarray*}
so that
\[ -c \:|\Phi^\pm|^2 \;\leq\; S \:\frac{d}{dr} |\Phi^\pm|^2 \;\leq\; 
        c \:|\Phi^\pm|^2 \]
where $c=2m+(2j+1)/\delta$. Dividing by $|\Phi^\pm|^2$ and integrating
gives, for $\delta < r < r^\prime < r_0$, or $r_1 < r < r^\prime$,
\begin{equation}
        -c \int_r^{r^\prime} S^{-1} \;\leq\; \left. \log |\Phi^\pm|^2 
        \:\right|_r^{r^\prime} \;\leq\; c \int_r^{r^\prime} S^{-1} \;\;\; .
        \label{eq:526}
\end{equation}
In the region $r_0<r<r_1$, (\ref{eq:517}) gives similarly
\[ S \:\frac{d}{dr} |\Phi^\pm|^2(r)
\;=\; \bra S \:\frac{d}{dr} \Phi^\pm,\: \Phi^\pm \ket
\:+\: \bra \Phi^\pm,\: S \:\frac{d}{dr} \Phi^\pm \ket \;=\; 0 \;\;\; , \]
since the square bracket in (\ref{eq:517}) is an anti-Hermitian matrix.
Thus $|\Phi^\pm|^2$ is constant in this region, so~(\ref{eq:526})
trivially holds for $r_0<r<r^\prime<r_1$.
Since $S^{-1}$ is integrable on the event horizons, (\ref{eq:526}) 
shows that $|\Phi^\pm|^2$ has finite boundary values on each side of 
the horizon, and these are non-zero unless if $\Phi^\pm$ vanishes 
identically on the corresponding region $(\delta, r_0)$, $(r_0, 
r_1)$, or $(r_1, \infty)$.

We now use (\ref{eq:519}) and (\ref{eq:520}) in the matching 
condition (\ref{eq:514}) to get for $j=0,1$,
\[ \Phi^\pm(r_j+\varepsilon) - \Phi^\pm(r_j-\varepsilon) \;=\;
o(1+|\Phi^\pm(r_j+\varepsilon)|) \spc,\;\;\; \varepsilon \rightarrow 0. \]
Since $|\Phi^\pm(r)|^2$ has 2-sided limits as $r_j$, we conclude that 
these limits must coincide at $r_j$; i.e.
\[ \lim_{0<\varepsilon \rightarrow 0} |\Phi^\pm(r_j+\varepsilon)|^2 \;=\;
\lim_{0<\varepsilon \rightarrow 0} |\Phi^\pm(r_j-\varepsilon)|^2 
\;\;\; . \]
Using (\ref{eq:526}) again, we conclude that the wave function 
vanishes on the entire interval $(\delta, \infty)$, if it is zero on 
$r_j$. This completes the proof since $\delta$ was arbitrary.
\QED
The final step is to use current conservation (cf.~\cite{FSY1})
\begin{equation}
\nabla_j \overline{\Psi} G^j \Psi \;=\; 0
        \label{eq:527}
\end{equation}
to study the decay of $\Phi^s_{jk \:\omega(n)}(r)$ at infinity, and 
to prove Part {\em{i)}} of Theorem~\ref{thm2}:
\begin{Thm} {\bf{(radial flux argument)}}
Either $\Psi^s_{jk\omega}$ vanishes identically, or the normalization 
condition (\ref{eq:525}) is violated.
\end{Thm}
{\Proof}
For simplicity, we again omit the indices $s$, $j$, $k$, and $\omega$.
Suppose that $\Psi \not \equiv 0$. For $r_1<r<R$ and $T>0$, let $V$ be 
the annulus outside the horizon $r$, given by $V=(0,T) \times (B_{2R} 
\setminus B_{2r})$. Using (\ref{eq:525}), we find
\begin{eqnarray*}
0 &=& \int_V \nabla_j (\overline{\Psi} G^j \Psi) \;\sqrt{|g|} \:d^4x \\
&=& \int_0^T dt \; r^2 \:S(r) \int_{S^2} (\overline{\Psi} \gamma^r 
\Psi)(t,r)
\:-\:\int_0^T dt \; R^2 \:S(R) \int_{S^2} (\overline{\Psi} \gamma^r 
\Psi)(t,R) \\
&&-\int_{2r}^{2R} ds \;s^2 \:S^{-1}(s) \int_{S^2} \left. 
(\overline{\Psi} \gamma^t \Psi)(t,r) \right|_{t=0}^{t=T} \;\;\;.
\end{eqnarray*}
Since the integrand is static, the last term vanishes, and we conclude that 
the radial flux is independent of the radius,
\begin{equation}
         r^2 \:S(r) \int_{S^2} (\overline{\Psi} \gamma^r \Psi)(r) \;=\;
        R^2 \:S(R) \int_{S^2} (\overline{\Psi} \gamma^r \Psi)(R) \;\;\; .
        \label{eq:528}
\end{equation}
Using (\ref{eq:519}) and (\ref{eq:520}), we have
\begin{equation}
r^2 \:S(r) \int_{S^2} (\overline{\Psi} \gamma^r \Psi)(r) \;=\; 
\int_{S^2} \Phi^*(r) \left( \begin{array}{cc} 0 & i \\ -i & 0 
\end{array} \right) \Phi(r) \;\;\; .
        \label{eq:529}
\end{equation}
The matching condition (\ref{eq:513}), expressed in terms of $\Phi$ 
gives
\[ \lim_{r_1 < r \rightarrow r_1} \left( \begin{array}{cc} 1 & i \\ i & 
-1 \end{array} \right) \Phi \;=\; 0 \spc . \]
Using this, we have from (\ref{eq:529})
\begin{eqnarray*}
\lim_{r_1 < r \rightarrow r_1} r^2 \:S(r) \int_{S^2} (\overline{\Psi} \gamma^r \Psi)(r)
&=& \lim_{r_1 < r \rightarrow r_1} \int_{S^2} \left[ \Phi^* 
\left( \begin{array}{cc} 1 & i \\ -i & 1 \end{array} \right)
\Phi \:-\: |\Phi|^2 \right] \\
&=& \lim_{r_1 < r \rightarrow r_1} \int_{S^2} \left[ \Phi^* 
\left( \begin{array}{cc} 1 & 0 \\ 0 & -1 \end{array} \right)
\left( \begin{array}{cc} 1 & i \\ i & -1 \end{array} \right) \Phi
\:-\: |\Phi|^2 \right] \\
&=&-\lim_{r_1 < r \rightarrow r_1}
\int_{S^2} |\Phi|^2 \;\neq\; 0 \;\;\; ,
\end{eqnarray*}
since $\Phi$ is finite and non-zero on the horizon $r_1$.

Now we consider the radial flux for large $R$. Since the flux is 
independent of $R$, we have from the last inequality
\begin{eqnarray*}
0 &<& \lim_{R \rightarrow \infty} |R^2 \:S(R) \int_{S^2} 
(\overline{\Psi} \gamma^r \Psi)(R)| \\
&=& \lim_{R \rightarrow \infty} |R^2 \:S^{-1}(R) \int_{S^2} 
(\overline{\Psi} \gamma^t \Psi)(R)| \;\;\;,
\end{eqnarray*}
because the metric is asymptotically Minkowskian.
Thus the integrand of our normalization integral
\[ (\Psi \:|\: \Psi)_\infty \;=\; \int_{2 r_1}^\infty dR \; R^2 
\:S^{-1}(R) \int_{S^2} (\overline{\Psi} \gamma^t \Psi)(R) \]
converges to a positive number, so that the normalization integral is 
infinite.
\QED
We have thus proved Part {\em{i)}} of Theorem~\ref{thm2}. For Part 
{\em{ii)}}, the case of an extreme RN background field, we use a quite 
different method; c.f.~\cite{FSY4}.

We remark that, using Chandrasekhar's separation method,
the results in this section can be extended to the 
axisymmetric case. Namely, for a quite general class of axisymmetric 
black-hole geometries, including the non-extreme Kerr-Newman 
solution, it is proved in~\cite{FKSY} that the Dirac 
equation admits no normalizable, time-periodic solutions.

\begin{footnotesize}
\hspace*{-.65cm} 
\begin{tabular}{lclcl}
\\
Felix Finster && Joel Smoller && Shing-Tung Yau \\
Max Planck Institute MIS && Mathematics Department && Mathematics 
Department \\
Inselstr.\ 22-26 && The University of Michigan && Harvard University \\
04103 Leipzig, Germany && Ann Arbor, MI 48109, USA
&& Cambridge, MA 02138, USA \\
\tt{Felix.Finster@mis.mpg.de} && \tt{smoller@umich.edu} &&
\tt{yau@math.harvard.edu}
\end{tabular}
\end{footnotesize}

\newpage

\end{document}